\documentstyle[12pt,aasms4]{article}

\begin{document}

\title{The REFLEX Galaxy Cluster Survey IV: The X-ray Luminosity 
Function\footnote{Based on observations at the European Southern 
Observatory La Silla, Chile}}
\author{H. B\"ohringer$^2$, C.A. Collins$^3$, L. Guzzo$^4$,
P. Schuecker$^2$, W. Voges$^2$, D.M. Neumann$^5$,
S. Schindler$^3$, G. Chincarini$^4$, S. De Grandi$^4$, 
R.G. Cruddace$^6$, A.C. Edge$^7$, T.H. Reiprich$^2$, P.~Shaver$^8$}

\altaffiltext{2}{Max-Planck-Institut f\"ur extraterrestrische Physik,
D-85740 Garching,Germany}
\altaffiltext{3}{Liverpool John Moores University, Liverpool,U.K.}
 
\altaffiltext{4}{Osservatorio Astronomico di Brera, Merate, Italy}
 
\altaffiltext{5}{CEA Saclay, Service d`Astrophysique, Gif-sur-Yvette, France}

\altaffiltext{6}{Naval Research Laboratory, Washington, USA}

\altaffiltext{7}{Physics Department, University of Durham, U.K.}

\altaffiltext{8}{European Southern Observatory, Garching, Germany}

\begin{abstract}
The X-ray galaxy cluster sample from the REFLEX Cluster Survey, which
covers the X-ray brightest galaxy clusters detected in the ROSAT
All-Sky Survey in the southern sky, is used to construct the X-ray
luminosity function of clusters in the local Universe.
With 452 clusters detected above an X-ray flux-limit 
of $3\cdot 10^{-12}$ erg s$^{-1}$ cm$^{-2}$ in 4.24 sr
of the sky, this sample is the most comprehensive X-ray cluster sample with a
well documented selection function, providing the best current 
census of the local X-ray galaxy cluster population. 
In this paper we discuss the construction of the luminosity
function, the effects of flux measurement
errors and of variations with sample region 
and we compare the results to those from previous surveys.
 
\end{abstract}


\section{Introduction}
Since the X-ray luminosity of galaxy clusters is closely related
to the cluster mass (Reiprich \& B\"ohringer 1999) and can be
measured for a large sample of galaxy clusters, the X-ray luminosity
function provides a good estimate of the mass function of galaxy clusters.
Therefore the X-ray luminosity function has been widely used as a census
of the galaxy cluster population in the Universe
(e.g. Picinotti et al. 1982,
Kowalski 1984, Gioia et al. 1984, Edge et al. 1990, Henry et al. 1992,
Burns et al. 1996, Ebeling et al. 1997, Collins et al. 1997, Rosati et al. 1998,  
Vikhlinin et al. 1998, De Grandi et al. 1999, Ledlow et al. 1999, Nichol et al. 2000,
Gioia et al. 2001). The close connection of cluster formation with the
evolution of the large-scale structure of the Universe makes the
cluster mass function  - and its observational substitute, the X-ray 
luminosity function - very important for  
the statistics of large-scale structure and for tests of 
cosmological models. The cluster luminosity function  
constrains in particular the normalization of the amplitude of
the primordial density fluctuation power spectrum on scales of
about 5 - 10 $h_{100}^{-1}$ Mpc (e.g. Henry \& Arnaud 1991, 
Bahcall \& Cen 1992, White et al. 1993), and
the evolution of the X-ray luminosity function provides a sensitive 
test of the mean density of the Universe
(e.g. Perrenod 1980, Oukbir \& Blanchard 1992, Eke et al. 1996, 
Viana \& Liddle 1996, Borgani et al. 1998). 

A precise measurement of this function had to await two
improvements: the availability of cluster samples large enough to reduce
the statistical scatter and effects of cosmic variance and homogeneous
enough to minimize uncertainties and corrections of selection effects. 
The ROSAT All-Sky Survey (RASS, Tr\"umper 1992, 1993), 
its improved processing (Voges et al. 1999), 
and a comprehensive optical follow-up observing program
provided the basis for such improvements. In this paper
we use the ROSAT-ESO Flux-Limited X-ray (REFLEX) cluster survey
(B\"ohringer et al. 1998, 2001 (paper I), Guzzo et al. 1999, 
Collins et al. 2000 (paper II), Schuecker et al. 2001 (paper III)) 
comprising 452 southern clusters in total and 449 with 
measured redshifts above a nominal X-ray flux limit of $3\cdot 10^{-12}$ 
erg s$^{-1}$ cm$^{-2}$ in the ROSAT band (0.1 - 2.4 keV)
to construct the X-ray luminosity function of galaxy clusters in 
the local Universe. 
Compared to previous cluster samples based on the RASS and used
for the construction of the X-ray luminosity function, the present
sample is more than a factor of two larger and features 
a well understood selection function. It provides a good measure
of the local luminosity function for studies of cluster evolution
by comparison with distant X-ray cluster samples (e.g. in Gioia et al. 2001). 

This paper is organized as follows. Section 2 gives a short
description of the REFLEX cluster sample.
The flux and luminosity determination is summarized in section 3.
In section 4 the X-ray luminosity function is derived and
comparison to previous results is made in section 5. 
Section 6 provides a summary. For the calculation of luminosities and 
volumina we use the cosmological parameters:
$H_0 = 50$ km s$^{-1}$ Mpc$^{-1}$,
$\Omega_0 = 1$ and $\Lambda = 0$. 

\section{The REFLEX cluster sample}

The construction of the REFLEX cluster sample is described
in detail by B\"ohringer et al. (2001).
The survey area covers the southern sky up to the declination 
$\delta = +2.5^o $, avoiding the band of the Milky Way 
($|b_{II}| \le 20^o $) and the regions of the Magellanic clouds. 
The total survey area is 13924 $\deg^2$ or 4.24 sr. 

The X-ray  detection of the clusters is based on the second
processing of the RASS (Voges et al. 1999), providing
54076 sources in the REFLEX area. All sources were reanalysed by means of 
the growth curve analysis (GCA) method (B\"ohringer et al. 2000) 
and the results are used to produce a flux-limited sample of RASS
sources with a nominal flux of $F_n \ge 3 \cdot 10^{-12}$
erg s$^{-1}$ cm$^{-2}$ (with $F_n$ as defined below). 
Cluster candidates were found using a machine based correlation of these
X-ray sources with galaxy density enhancements in 
the COSMOS optical data base (derived from digital scans of the
UK Schmidt survey plates by COSMOS at the Royal Observatory Edinburgh,
MacGillivray \& Stobie 1984). 
The resulting candidate list was carefully screened based on
X-ray and optical information, literature data, and results
from the optical follow-up observation program. The selection 
process was designed to provide a completeness in the final cluster catalogue 
in excess of 90\% and several further completeness tests support this claim. 
The final cluster sample includes
452 clusters and there are three objects left in the list with uncertain
identifications and redshifts. These three objects are excluded 
from the further analysis. The sample has already been used 
to analyse the statistics of the spatial cluster distribution 
by the two-point correlation function (Collins et al. 2000) and  
by the density fluctuation power spectrum (Schuecker et al. 2001).

\section{Luminosity and survey volume determination}
 
The X-ray luminosities of the REFLEX clusters are determined from the
count rate measurements provided by the GCA
(B\"ohringer et al. 2000). For the first analysis these count rates
are not corrected by means of a model estimate of the total flux.
Such modifications are discussed in a second step.
To determine the cluster X-ray luminosity we convert the measured
count rate into a ``nominal'' X-ray flux for the ROSAT band (0.1 to
2.4 keV), $F_n$, by assuming a Raymond-Smith type spectrum 
(Raymond \& Smith 1977) for a temperature of 5 keV, a metallicity of 
0.3 of the solar value (Anders \& Grevesse 1989), a redshift of zero,
and an interstellar hydrogen column density as found for the line-of-sight 
in the compilation by Dickey \& Lockman (1990) 
as given within EXSAS (Zimmermann et al. 1994).
The value of $F_n$ is used to make the flux cut independent of 
any redshift information (since this information is not available
for all objects at the start of the survey).
With the redshift value at hand, the X-ray flux is redetermined ($F_x$)
with an improved spectral model, where the temperature is now 
estimated (iteratively) from the preliminarily derived X-ray luminosity
and the luminosity-temperature relation derived by Markevitch (1998).
The redshift of the spectrum is now taken into account, which corresponds
to a $k$-correction in optical astronomy with $k(T,z,N_H)$ and
provides luminosities for the cluster rest frame energy band 0.1 to 2.4 keV. 

For the construction of the luminosity function of a flux-limited
sample the survey volume, $V_{max}$, as a function of X-ray luminosity 
has to be known. The survey volume is given by the volume of the 
cone defined by the survey area and the luminosity distance at which 
a cluster with a given luminosity is just observed at the flux limit. 
Since we have used the flux parameter $F_n$ for the flux cut 
$F_{n~lim}$ and since we have calculated the luminosity, $L_x$, 
iteratively from $F_n$, we have to reverse these steps, where the 
limiting luminosity distance, $D_{L~lim}$, is now iteratively calculated 
from $F_{n~lim}$ and $L_x$ involving the two steps:
$ corr = F_{x~lim}/F_{n~lim} = f(L_x,D_{L~lim}) $  and 
$ D_{L~lim}^2 = {L_x \over 4\pi~~ F_{n~lim}~~ corr(L_x,D_{L~lim})~~
k(T,z,N_H)} $.
 
These equations establish a unique relation between $L_x$ and $V_{max}$
for given $F_{n~lim}$. The two correction factors are small with
typical values quoted by B\"ohringer et al. (2000).
(Note, that no new flux cut has been introduced after the correction
of the flux values. Therefore the sample does not change and 
the selection volume depends on $F_{n~lim}$).

The second correction applied in the $V_{max}$ calculation concerns
the sensitivity function derived in paper I providing the sky
coverage as a function of flux (B\"ohringer et al. 2001, Figs. 22
and 23). The sensitivity function is
defined by two limiting parameters, the flux limit, $F_{n~lim}$, and the minimum
number of photons required for a safe detection and flux measurement.
We use a soft coding of the photon number
cut, such that the effect of using different cut values can easily be
explored. For a minimum value of 10 photons, for example, the nominal 
flux limit is reached in 97\% of the REFLEX area, while for a value of 
30 photons this fraction 78\%. For the remaining part of 
the sky with higher flux limit the corresponding 
survey volume has to be reduced accordingly. The large
sample size of REFLEX allows us to be selective and to use the 
very safe, higher cut of 30 photons for the standard derivation of the 
luminosity function. 

As shown in paper I, a complete removal of the photon 
number cut leads to an estimated deficit of only about $14 (\pm 7)$  
clusters (3.8\%). Thus the assumption of a homogeneous selection function
without source count limit and inclusion of all clusters
leads to results insignificantly different from the results
obtained with the conservative approach below.
For a minimal photon number of 30, the sample
contains 423 clusters with redshifts. For luminosities lower than
$L_x = 10^{42}$ erg s$^{-1}$ the counterparts to the extended X-ray 
sources very often appear as single elliptical
galaxies with no optically bright companions and therefore these
objects may be incompletely represented in our sample. 
We therefore exclude three objects with lower luminosity 
from the parametric fit to the luminosity function
described below.

\section{Results}

\begin{figure}
\plotone{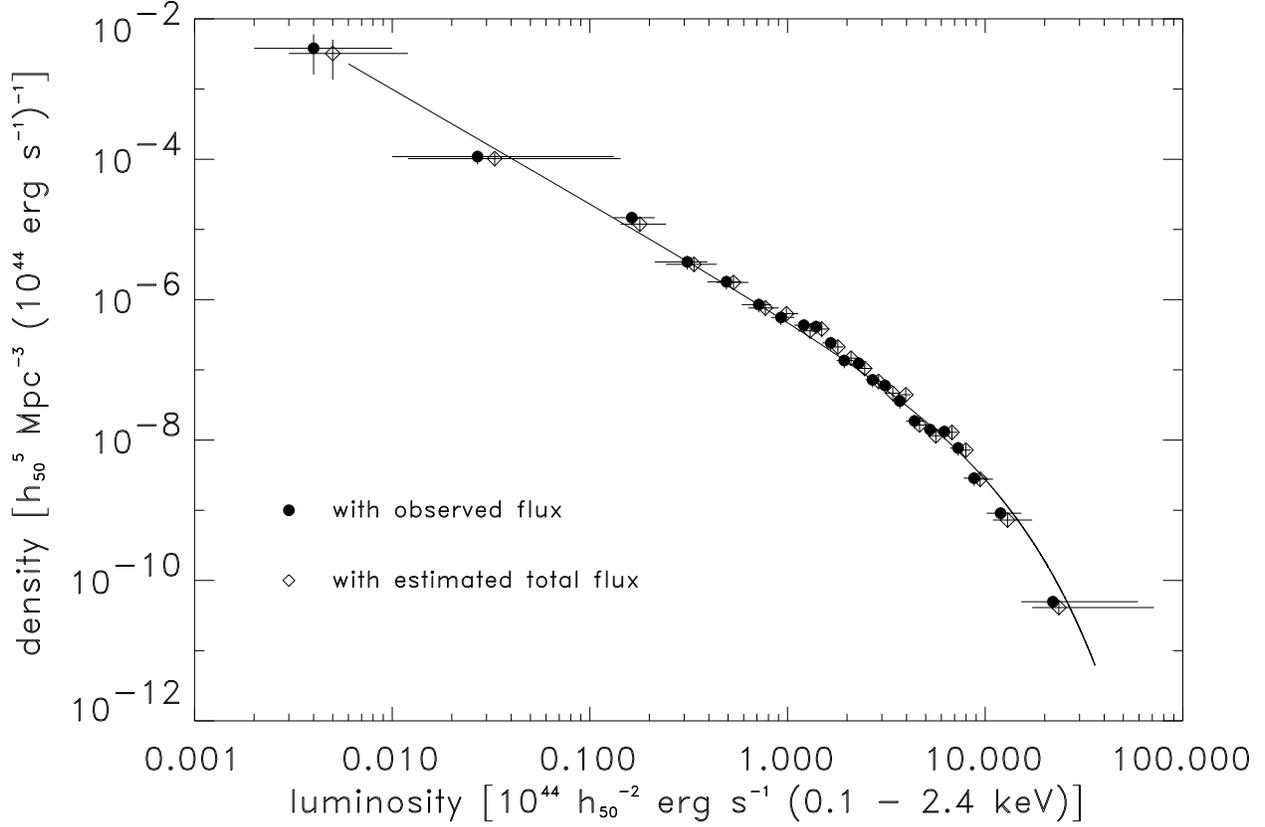}
\caption{X-ray luminosity function for the REFLEX sample. The solid
data points give the results for the detected luminosities while the
open diamonds indicate the results including a model dependent correction
for the missing flux. The data points are plotted at the density weighted mean luminosity
per bin. The line gives the maximum likelihood
fit including the correction for missing flux and the individual uncertainties
in the flux measurement.} 
\end{figure}

The binned luminosity function calculated for this sample is shown in
Fig. 1 (with 20 clusters per bin). The 3 objects
with the lowest luminosity are grouped here into the first bin.  
The calculation for each bin is performed using the formula

$$ n(L)  = {1\over \Delta L} \sum_{i=1}^{N}~ {1 \over V_{max}(L_i)} $$ 

where the sum is over all $N$ clusters falling into the luminosity interval
of the bin. Note, that with this approach we determine the mean luminosity
function in the survey volume averaging over large-scale structure density
fluctuations. Also, no major redshift dependence of the cluster density has
been detected.
An alternative determination of a density independent luminosity
function and the redshift dependence of the cluster
density is planned for a future paper. 
The error bars shown are Poissonian errors based on
the number of clusters per bins. The function shown in Fig. 1 by
solid data points concerns the observed fluxes only. To explore
the effect of the flux missed by the GCA algorithm in the outskirts
of the clusters we correct the
fluxes and luminosities based on a self-similar cluster model as 
described in B\"ohringer et al. (2000): a $\beta$-model 
(Cavaliere \& Fusco-Femiano 1976) with a $\beta$-value of 2/3,
a core radius that scales with mass, 
and an assumed extent of the X-ray halo out to 
12 times the core radius. The correction procedure has been
successfully tested by simulations based on the same cluster model.
The resulting corrected luminosity function is also shown in Fig. 1.
As expected from the typical mean correction factor of about 8\%  
(B\"ohringer et al. 2000) the main effect is a shift of the curve
to higher luminosity of this order. A larger shift is only 
observed for the lowest 
redshift bins (for the groups with extended, low surface brightness
emission).

As a first consistency check we compared luminosity functions
derived for flux limits of $3\cdot 10^{-12}$ and 
$5\cdot 10^{-12}$ erg s$^{-1}$ cm$^{-2}$ finding excellent agreement.
This indicates that there is no significant 
incompleteness effect at low fluxes.
We also derived an unbinned, parametric representation
of the luminosity function by means of a maximum likelihood 
fit (by the method described by e.g. Murdoch et al. 1973 and by an alternative 
Poisson formulation of the ML method, with both giving identical results) 
of a Schechter function 

$$ n(L) dL~ =~ n_0~\exp (-{L \over L_{\star}})~ 
\left({L \over L_{\star}}\right)^{-\alpha}~ {dL \over L_{\star}} $$    

(Schechter 1976) over the luminosity range
$10^{42}$ erg s$^{-1}$ to infinity.  
The resulting best fit parameters are given in Table 1. The 
normalization parameter $n_0$ is derived from the requirement that
the total number obtained by integration of $n(L)$ equals the
observed number of clusters. The constraints obtained from the 
maximum likelihood analysis for the shape parameters are shown in
Fig. 2. We have also performed a $\chi^2$ fit to the binned data
to test for the quality of the fit and obtained a $\chi^2$ value 
of 18 (for 19 dof, 22 bins), 39 (for 40 dof, 43 bins), and 64 (for 
68 dof, 71 bins). Thus the Schechter function provides a 
good represetation of the data within the current uncertainty limits. 
The $\chi^2$ method is also used
for the error estimation for $n_0$. 

\begin{deluxetable}{llll}
\tablecolumns{4}
\tablewidth{0pc}
\scriptsize
\tablecaption{Results of the fitting of a Schechter function to 
REFLEX X-ray luminosity function and results from previous work}
\tablehead{
\colhead{sample} & \colhead{$L_{\star}^{a)}$} & \colhead{$\alpha$}& \colhead{$n_0^{b)}$}}
\startdata
REFLEX uncorrected  & $6.26{+0.6\choose -0.53}^{c)}$ & 1.63($\pm 0.06$)& $1.75{+0.5\choose -0.4}\cdot 10^{-7}~^{d)}$ \nl 
corr. for miss. flux & $6.79{+0.6\choose -0.55}$ & 1.63($\pm 0.06$) & $1.80{+0.5\choose -0.4}\cdot 10^{-7}$ \nl
corr. for flux error & $6.00{+0.6\choose -0.5}$ & 1.63($\pm 0.06$) & $1.58{+0.5\choose -0.4}\cdot 10^{-7}$ \nl
corr. for both effects & $6.47{+0.6\choose -0.53}$ & 1.63($\pm 0.06$) &$1.68{+0.5\choose -0.4}\cdot 10^{-7}$ \nl
High flux sample uncorr.$^{\ast}$ & 6.85($\pm 0.7$) & 1.68($\pm 0.07$) & $1.5{+0.6\choose -0.5}\cdot 10^{-7}$ \nl
BCS  & $9.1{+2.0\choose -1.5}$ & $1.85(\pm 0.09)$ & $7.74{+0.76\choose -0.70}\cdot 10^{-8}$\\
Bright RASS1 & $6.08 {+1.1\choose -0.9}$ & 1.52($\pm 0.11$)& $2.53(\pm 0.23)\cdot 10^{-7}$\\
Ledlow and others & $8.78(\pm0.62)$ & $1.77(\pm 0.01)$ & $7.9(\pm 0.38) \cdot 10^{-8}$\\
\enddata
\tablenotetext{}{$\ast$) sample with a flux-limit of $5\cdot 10^{-12}$ 
erg s$^{-1}$ cm$^{-2}$}
\tablenotetext{}{$^{a)} 
L_{\star}$ is in units of $10^{44} h_{50}^{-2}$ erg s$^{-1}$ cm$^{-2}$ 
in the 0.1 to 2.4 keV band}
\tablenotetext{}{$^{b)} n_o$ is in units of $h_{50}^3$ Mpc$^{-3}$}  
\tablenotetext{}{$^{c)}$
errors are quoted for 68\% limits}
\tablenotetext{}{$^{d)}$ the errors quoted for the normalization for the REFLEX
samples were evaluated by a $\chi^2$ method with two free parameters
($L_{\star}$, $\alpha $), which differes from the approach for the other samples.
For one free parameter the error reduces to $\pm 0.1$ -- $\pm 0.2$ and to
$\pm 0.08$ for fixed ($L_{\star}$, $\alpha$) and Poissonian errors.}
\end{deluxetable}

In the next step of the analysis we consider the effect of the
uncertainties in the flux measurement on the results. 
For this purpose we extend the 
maximum likelihood approach of Murdoch et al. to include the effect
of errors on the expected luminosity distribution as well as
on the uncertainty of the survey volume (the Eddington (1940) bias). 
The $\log$-likelihood, ${\cal L}$, is then given by:

$$ {\cal L}~ = \sum_i ~-\ln K(\sigma _i)~ + ~\ln 
\int_{L_{min}}^{\infty}~{V_{max}(L^{\prime})~
n(L^{\prime})~ G(L_i,L^{\prime},\sigma _i)~ dL^{\prime}} $$

$$ {\rm with} ~~~ K(\sigma _i)~ =~ \int_{L_{min}}^{\infty}~ dL~ 
\int_{L_{min}}^{\infty}~ dL^\prime~ 
V_{max}(L^\prime)~ n(L^\prime)~ G(L,L^\prime,\sigma _i)$$

where $G(L,L^\prime,\sigma _i)$ is a normalized Gaussian function
containing the photon noise error of the flux measurement, 
$\sigma _i$, for each cluster and the sum is over all clusters in 
the sample. While the previous correction for missing flux leads
essentially to an increase in $L_{\star}$ of about 8\%, the inclusion
of the flux errors results in a decrease by about 4\%. Both effects
are relatively small, yielding overlapping parameter constraints
(Fig. 2).

\begin{figure}
\plotone{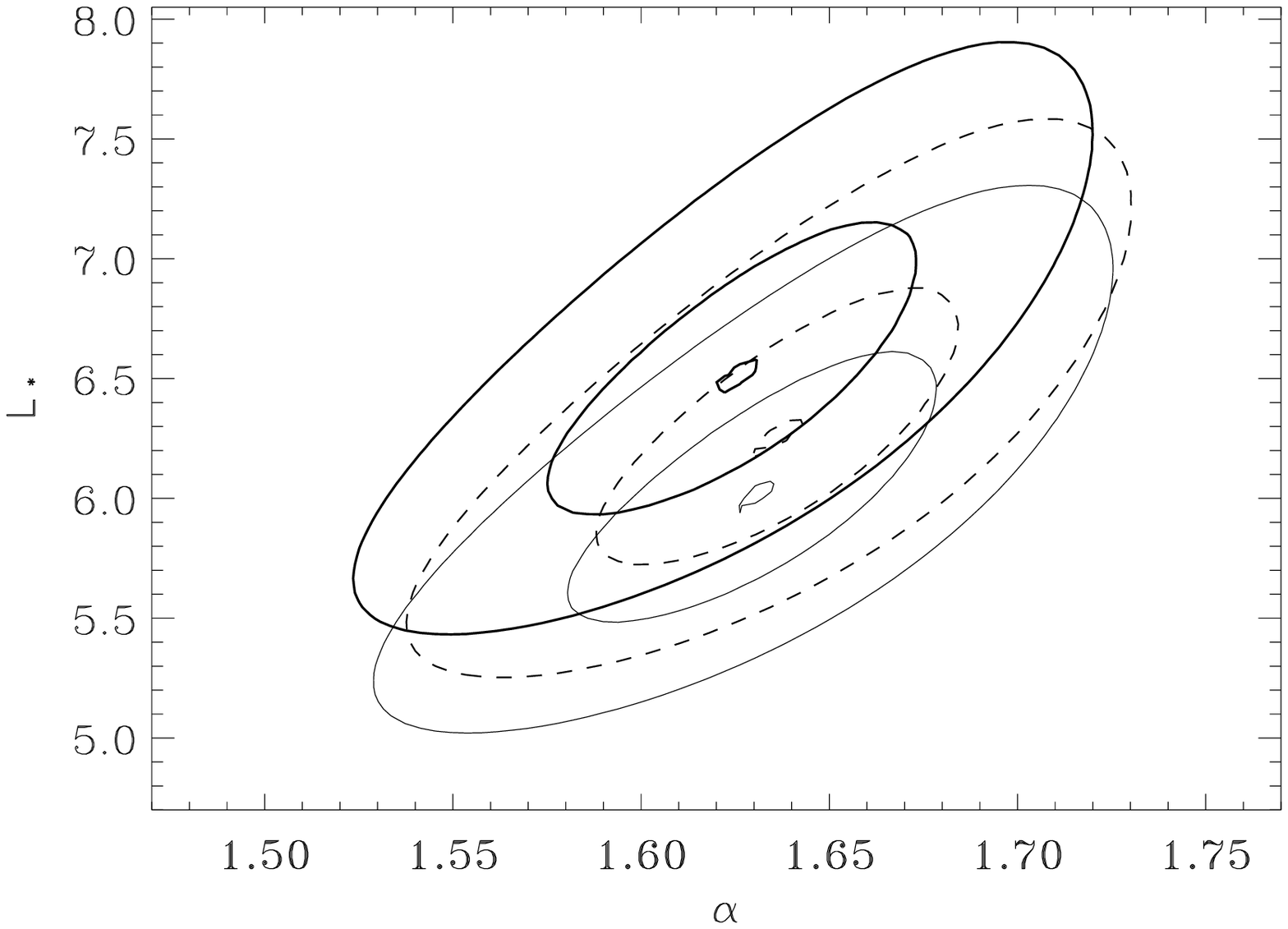}
\caption{Constraints on the parameters of the shape of the Schechter function
derived from maximum likelihood fits to the total sample including
corrections for missing flux and flux errors (solid lines), uncorrected
for missing flux (thin lines), no correction for missing flux and flux
error (broken line). 
The contour lines encircle the best fitting value and indicate the 
$1\sigma$- and $2\sigma$-limits, respectively.}
\end{figure}

As another test of the stability of the results we compare in Fig. 3
and Table 1 the results obtained for the luminosity function,
if the REFLEX sample is split into the part above and 
below the galactic disk. There is good agreement within the
error bars at intermediate luminosities where most of the 
clusters were found. The deviations at low luminosities are 
consistent with the cosmic variance estimated for the respective
survey volume (approximated to be spherical) and the power spectrum
determined for the REFLEX cluster distribution (Schuecker et al.
2001). Details of this calculation will be given in a further 
publication in this series. The differences at the highest
luminosity are due to small number statistics.

\begin{figure}
\plotone{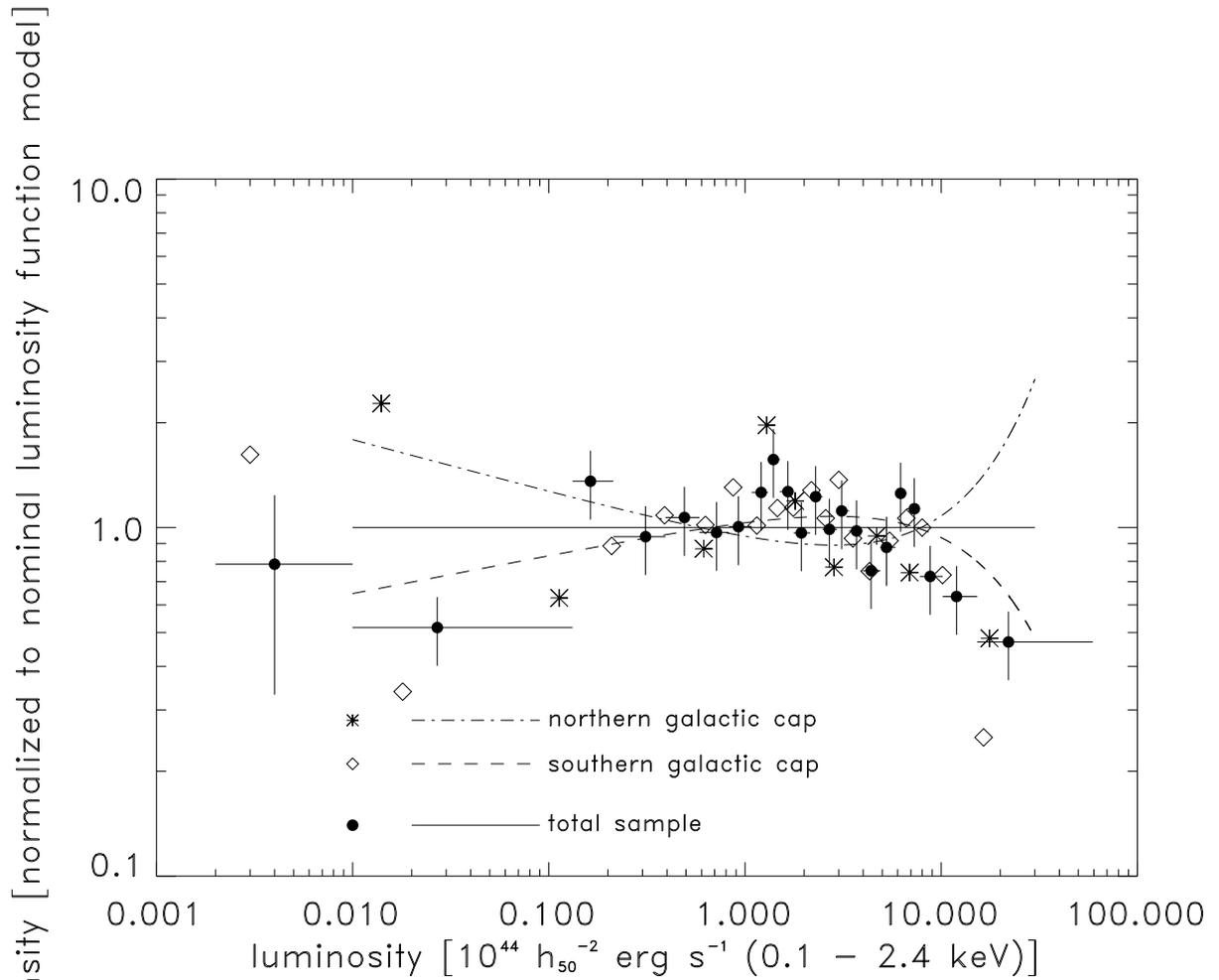}
\caption{Comparison of the X-ray luminosity function derived for the
subsamples in the southern and northern galactic caps with the result
of the total REFLEX sample. All results are normalized by the luminosity
function for the total sample (solid line). All functions are given in
the observed, uncorrected form.}
\end{figure}

\section{Comparison to previous results}

In Fig. 4 we compare the results for REFLEX with the  
largest previous samples from the first processing 
of the RASS, our RASS1 Bright Sample in the South (De Grandi et al. 1999), 
BCS in the North (Ebeling et al. 1997) and the work by Ledlow et al. (1999)
based on X-ray detections of Abell clusters.
We note that even though the results agree within the combined
individual errors in this binned representation there are global 
differences. At low luminosities the differences are approximately 
within the expected cosmic variance (e.g. Fig. 3).
The low  density at low luminosities in the Ledlow et al. sample
is due to the fact that the Abell catalogue does not well sample the
X-ray emitting galaxy groups. 
At medium luminosity, where the data sets are
most accurate, the result for the RASS Bright Sample are systematically higher
and the BCS and Ledlow et al. samples are lower by about 20-30\%.
The De Grandi et al. results predict a cluster density 
for the most interesting part of the luminosity function
which is about 50\% higher than that of the BCS
(noted also by Gioia et al. 2001). Therefore this comparison shows
where an improvement in the precision of the luminosity
function is achieved -- in particular as reference for the local Universe 
in the study of cluster evolution.

\begin{figure}
\plotone{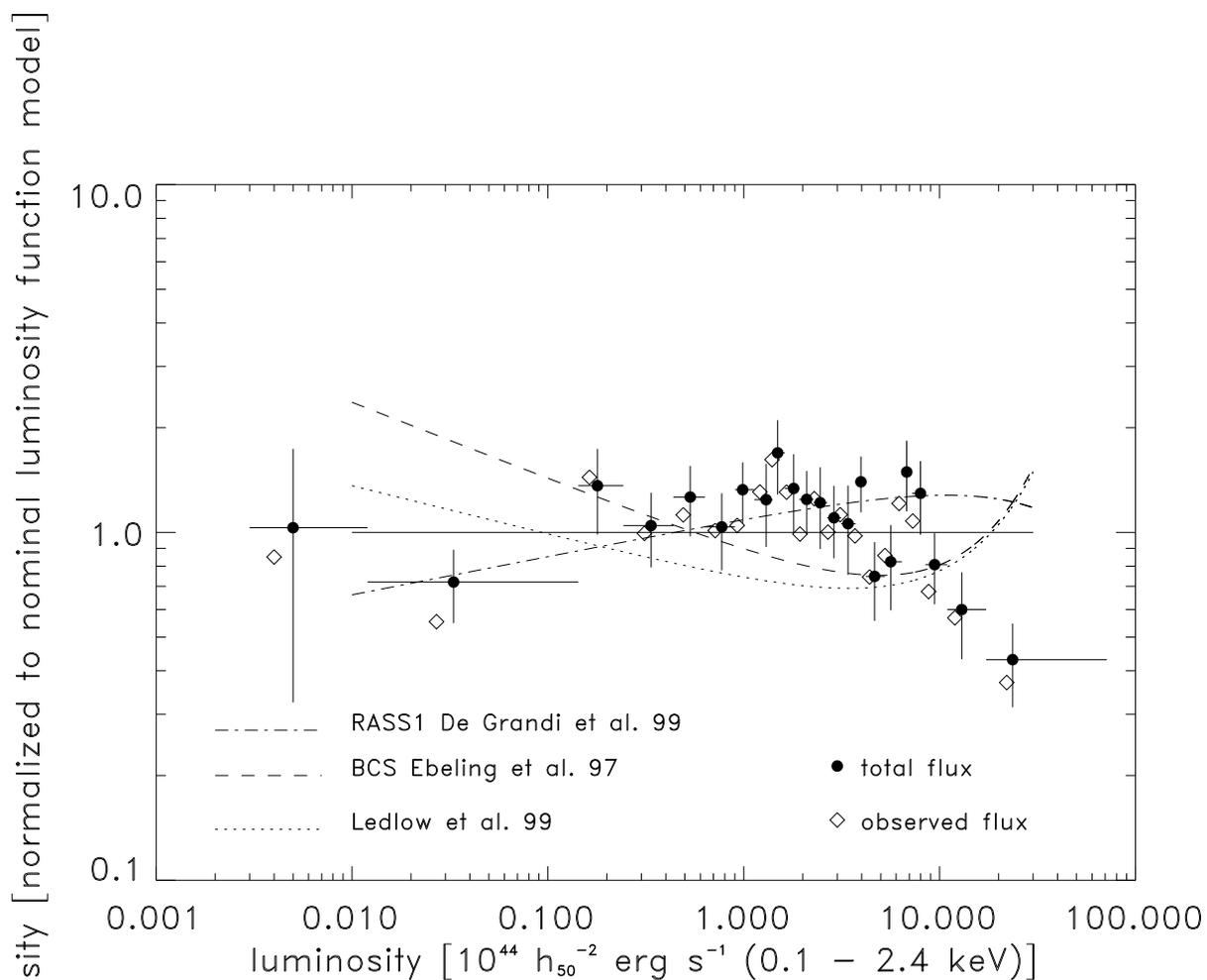}
\caption{Comparison of the REFLEX X-ray luminosity function to those
of the BCS (Ebeling et al. 1997), the Bright RASS1 sample 
(De Grandi et al. 1999), and X-ray detected Abell clusters (Ledlow et al. 1999).
All results are normalized by the luminosity
function for the REFLEX sample (solid line). The REFLEX function is used in
the form corrected for missing flux but uncorrected for the flux
errors, to conform with the treatment of the other surveys. 
The REFLEX data points without missing flux correction are also shown.}
\end{figure}

\section{Summary and conclusions}

The REFLEX sample has allowed us to determine the X-ray luminosity 
function with an accuracy of for example better than 20\% 
for 22 independent data bins
over three orders of magnitude in luminosity
(better than 12\% for 8 independent $e$-folding intervals). 
This accuracy will 
provide the basis for a precise comparison with 
luminosity functions determined for
high redshift samples in search for evolutionary effects.
The size of the REFLEX sample has also allowed us to determine
the luminosity function of subsamples for different flux-limits
and different survey regions to demonstrate the stability of the result.
In our following work we will use the X-ray luminosity function
derived here to obtain constraints on cosmological models.

\acknowledgments
We thank Joachim Tr\"umper and the ROSAT team providing the RASS data
fields and the EXSAS software as well as H.T. Mac Gillivray, Daryl Yentis,
and the COSMOS team for the digitized optical data.
P.S. acknowledge the support by the Verbundforschung under the grant
No.\,50\,OR\,9708\,35, H.B. the Verbundforschung under the grand
No.\,50\,OR\,93065.

\end{document}